\documentclass[preprint]{aastex}
\usepackage{epsfig}
\newcommand{\vi}{\mbox{$V\!-\!I$}}

\slugcomment{Accepted for publication in the Astronomical Journal, July 2001} 

\shorttitle{Cluster with a peculiar BCG}
\shortauthors{K.A. Williams}

\begin{document}

\title{Serendipitous discovery of a cluster of galaxies with a
peculiar central galaxy\footnote{Based
on data obtained at the W. M. Keck Observatory, which is operated as a
scientific partnership among the California Institute of Technology,
the University of California, and the National Aeronautics and Space
Administration.  The Observatory was made possible by the generous
financial support of the W. M. Keck Foundation.}}

\author{Kurtis A. Williams}
\affil{Department of Astronomy and Astrophysics}
\affil{University of California, Santa Cruz, CA, 95064}
\email{williams@ucolick.org}

\begin{abstract}
We report the serendipitous discovery of a cluster of
galaxies at $z=0.369$.
Thirty-eight candidate members were identified based on rough broad-band
photometric redshifts, and three members were confirmed
spectroscopically.  The brightest cluster galaxy (BCG) is exceptionally
blue, with \bv$=0.12$ and \vi$=1.02$. 
The surface-brightness profile of the BCG follows an $r^{1/4}$-law
profile out to 3\arcsec ~in all three bands. The effective
radius is significantly smaller in bluer bandpasses, resulting in a
blue core and a color gradient opposite to the metallicity-induced
color gradient observed in typical elliptical galaxies.  Beyond
3\arcsec ~an extended envelope of emission in excess of the $r^{1/4}$-law
profile is observed, the position angle of which 
coincides with the major axis of the galaxy cluster.  The spectrum of
the BCG contains strong Balmer absorption, a minimal 4000\AA ~break,
and a broad \ion{Mg}{2} emission line, suggesting that the galaxy has
undergone recent star formation and may
harbor a central AGN.  The presence of numerous nearby
bright stars makes this cluster an interesting target for next-generation 
adaptive optics using natural guide stars.
\end{abstract}

\keywords{galaxies: active --- galaxies: elliptical and lenticular, cD
--- galaxies: clusters: general}

\section{Introduction}
The formation and evolution of clusters of galaxies and their
constituent members is one of the most active areas of research in
extragalactic astronomy.  Numerous large-scale surveys at
x-ray \citep[e.g.][]{romer00,scharf97} and optical
\citep[e.g.][]{gonzalez01,scodeggio99} wavelengths are rapidly 
increasing the number of known galaxy clusters at intermediate
redshifts.  

The central regions of many of these clusters, and of many clusters at
low redshifts, contain a cD galaxy -- an elliptical galaxy surrounded by
an extended envelope of diffuse emission
\citep{schombert86,schombert88}.  Studies of these galaxies have
indicated that their formation involves complex and uncertain
processes. It is believed by many groups 
that these luminous central galaxies were created by hierarchical
merging in the distant past \citep{merritt85,dubinski98},
as the velocity dispersions in relaxed galaxy clusters are too high to
permit merging of galaxies. Still, other formation scenarios have
not yet been ruled out \citep[see discussion in][]{garijo97}. The
formation mechanism for the cD envelope also remains controversial
\citep[e.g.][]{mclaughlin93}. 

We present the serendipitous discovery of a cluster
of galaxies at an intermediate redshift, $z=0.369$.  The brightest
cluster galaxy shows many of the properties of a cD galaxy, including a
deviation from an otherwise smooth $r^{1/4}$ luminosity profile,
a high luminosity, and a surrounding cluster of galaxies. However,  
closer inspection reveals many peculiarities, including a strong color
gradient, a potential AGN, and evidence of recent star formation. This unusual
combination of traits makes this 
cluster of galaxies a candidate for the study of cD galaxy
formation.  

Here we briefly discuss the observations (\S 2) before detailing the
candidate selection and photometric analysis (\S 3), including a
discussion of the surface-brightness profile of the BCG.  In \S 4 we
discuss the spectroscopic results, focusing on the central
galaxy (\S 4.1).  We conclude with a discussion of the central galaxy,
including speculation on the origins of its unusual spectral and
photometric properties, as well as the possibilities for future study (\S5).

In this paper, we assume $H_0 = 70$ km s$^{-1}$ Mpc$^{-1}$, $\Omega_M
= 0.3$, and $\Omega_\Lambda = 0.7$.  With these parameters, $1\arcsec$
corresponds to a physical distance of 8.8 kpc at the cluster
redshift. 

\section{Observations}
\subsection{Discovery images}
The cluster of galaxies was discovered in images of 
the star cluster M67 taken using the imaging mode of the LRIS imaging
spectrometer 
\citep{oke95} on Keck II on 1 February 1997. These observations are
described in detail in 
\citet{williams00}.  For this analysis, we select only
the small subset of images wholly containing the bright central
galaxy.  Images taken with each filter were shifted and co-added,
resulting  in total exposure times of 1200s in $B$, 600s in  
$V$, and 360s in $I$. The FWHMs for each band, determined by
measurements of PSF stars with DAOPHOT \citep{stetson87}, were
$0\farcs 78$ in $B$, $0\farcs 83$ in $V$, and $0\farcs 65$ in $I$.
The galaxy  cluster is located on the northern edge of the
frames.  The section of the image containing the cluster is shown in
Fig.~\ref{clusterimage}. A search of the NASA/IPAC Extragalactic
Database finds no objects listed within a 2\arcmin ~radius of the
brightest cluster galaxy (BCG). 

The images were reduced and flat-fielded via standard methods.
Calibration was performed in each filter using local calibration
stars from the M67 photometry of \citet{mont93}.  Source
detection, photometric analysis, and shape analysis were performed
using the SExtractor data analysis program \citep{bertin96}.
Objects are detected to $V\approx 26.5$.  
SExtractor assigns a stellarity parameter to each object with a value
between 0 and 1, with 0 indicating an extended source, 1 indicating a
point source, and intermediate numbers indicating a degree of
uncertainty about the classification.  The measured stellarity
parameters of objects in  these images reliably separate point sources
from extended sources for $V\lesssim 24$, corresponding to an absolute
magnitude of $M_V\approx -18.6$ for an E/S0 galaxy at the cluster
redshift \citep{coleman80}. We consider all objects brighter than 
$V=24$ and having shape parameters $\leq 0.2$ to be
galaxies. All other objects, being either stellar or ambiguous
classifications, were excluded from further analysis.

\subsection{Spectroscopic observations}
Spectra were taken in the same field of view as the images during a
follow-up run on 26 and 27 February 2000 at Keck I.  LRIS was used in
spectroscopic mode with the 600 mm$^{-1}$ grating and a multi-slit mask with
slitlet widths of $1\farcs 1$.  Seeing was highly variable, averaging
around 1 arcsecond.  The spectroscopic setup was optimized for study
of faint white dwarfs and planetary nebulae, and the covered wavelength
range ($\sim$4000\AA ~to $\sim$6000\AA) is less than ideal
for analysis of the cluster members.  Spectroscopic resolution is
measured to be $\sim 5$\AA, and the total exposure time is 3600s.

The spectra were bias-corrected
 and flat-fielded via standard methods in
IRAF.  Relative flux calibration was performed using spectra of the
spectrophotometric standard star \objectname{Feige 66}, although the
absolute calibration is considered unreliable due to the use of a
 different slit width ($0\,\farcs 9$) and the highly variable
 seeing. The relative calibrations are also considered unreliable for
 $\lambda \lesssim 4000 \mbox{\AA}$ due to
the rapid decline in instrumental response and for $\lambda \geq 5965
 \mbox{\AA}$, since the calibration spectrum ends at this point.
No corrections were made for atmospheric extinction.

\section{Candidate Galaxy selection and photometry}

Galaxy photometry was performed using the SExtractor analyses of the
images.  It has 
been noted that SExtractor systematically underestimates the total
flux of an object \citep{bertin96}, and we observed that the various
magnitudes calculated by SExtractor (e.g. isophotal magnitudes, corrected
isophotal magnitudes, and ``best'' magnitudes) did not agree, and that
the disagreement was worse for fainter objects.  In order to estimate
the true total magnitudes for each galaxy, we created several populations of
artificial galaxies in each band using standard IRAF routines.  

\begin{figure}
\epsscale{1}
\plotone{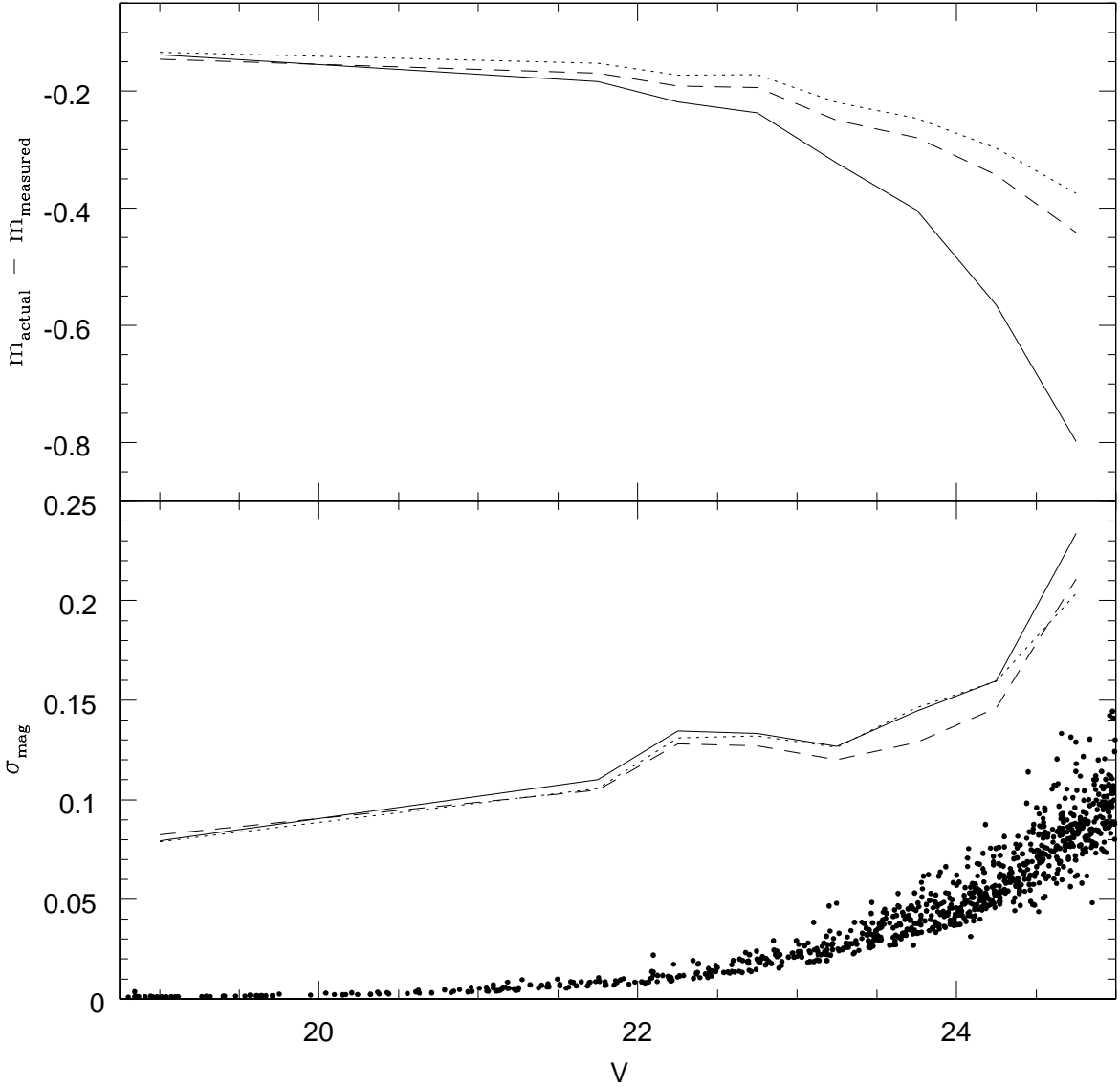}
\caption{Errors in SExtractor
$V$-band photometry for a population of artificial galaxies.  The top panel
shows the offset between actual and measured magnitudes as a function
of magnitude for three types of output magnitudes: isophotal
magnitudes (solid line), corrected isophotal magnitudes (dashed line),
and ``best'' magnitudes (dotted line).
The lower panel plots the standard deviation in
measured magnitudes about the mean offset as a function of
magnitude for each type of magnitude measurement (lines) and the
SExtractor-determined errors for each detected galaxy (points). \label{photerrs}} 
\end{figure}

The artificial 
galaxy luminosities are spaced randomly within 0.5-magnitude-wide
bins, are convolved with the appropriate
seeing for each band, and are placed
randomly in the region of the images containing the cluster.  25\% of
the artificial population are exponential disk galaxies, while the
remainder have $r^{1/4}$-law luminosity profiles.  The scale radii and
effective radii are set to 2.5 pixels, or approximately 5 kpc at the cluster
distance. The artificial galaxies are detected and analyzed using SExtractor,
and the mean magnitude offset and standard deviation about the offset
are calculated for each bin in each band.  
The results for the $V$-band are shown in Fig.~\ref{photerrs}; other
bands are omitted from the figure for clarity.

The suspected systematic offset between SExtractor
magnitudes and known artificial galaxy magnitudes is confirmed, with
the SExtractor magnitudes at least 0.1 magnitudes fainter than the
actual magnitudes.  This discrepancy increases toward fainter
magnitudes, and the discrepancy differs with the observed band.
Our measured systematic offsets are consistant with the offsets 
discussed in \citet{bertin96}.  The photometric measurements for each
galaxy are corrected for the missing flux using a
cubic spline interpolation of the measured offsets.  After correction,
the various calculated magnitudes for a given galaxy agree to within
the photometric errors.

The standard deviations of the artificial-galaxy magnitudes about the
mean offset are much larger than the SExtractor-calculated errors.
The source of this discrepancy is unclear. We
therefore interpolate between the standard deviations measured in each 
magnitude bin to find the error for each photometric measurement.  Errors
for colors are determined by adding the errors from the two
bandpasses in quadrature. The standard
deviations from the galaxy simulations for the $V$-band, as well as
the SExtractor-calculated $V$-band errors,  are plotted as an example
in the lower panel of Fig.~\ref{photerrs}.  

\begin{figure}
\epsscale{0.58}
\plotone{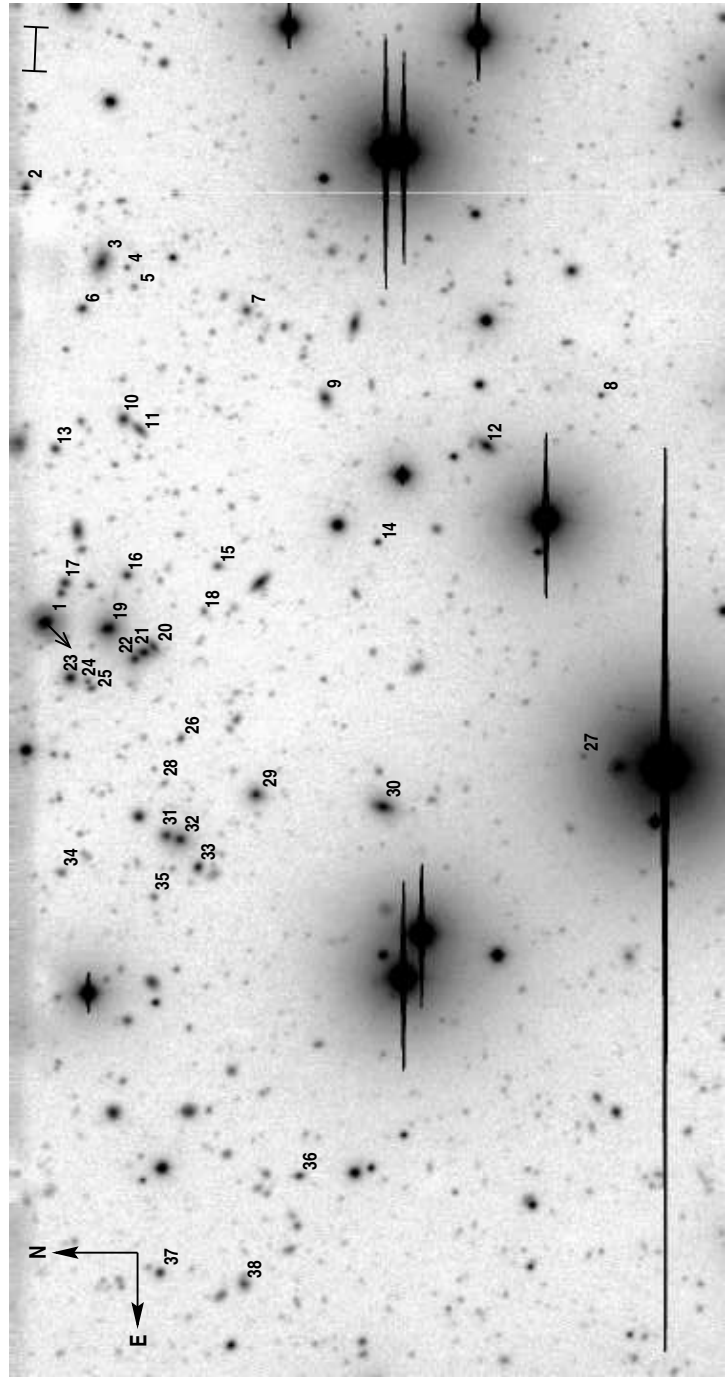}
\caption{Combined $B,\,V,\,I$ greyscale image of the
cluster of galaxies.  Candidate cluster members are indicated by
object number.  The scale bar 
in the upper right-hand corner represents 10 arcseconds.  The position
angle of the outer envelope of the central galaxy is indicated by the
dashed arrow. \label{clusterimage}}
\end{figure}

\begin{figure}
\plotone{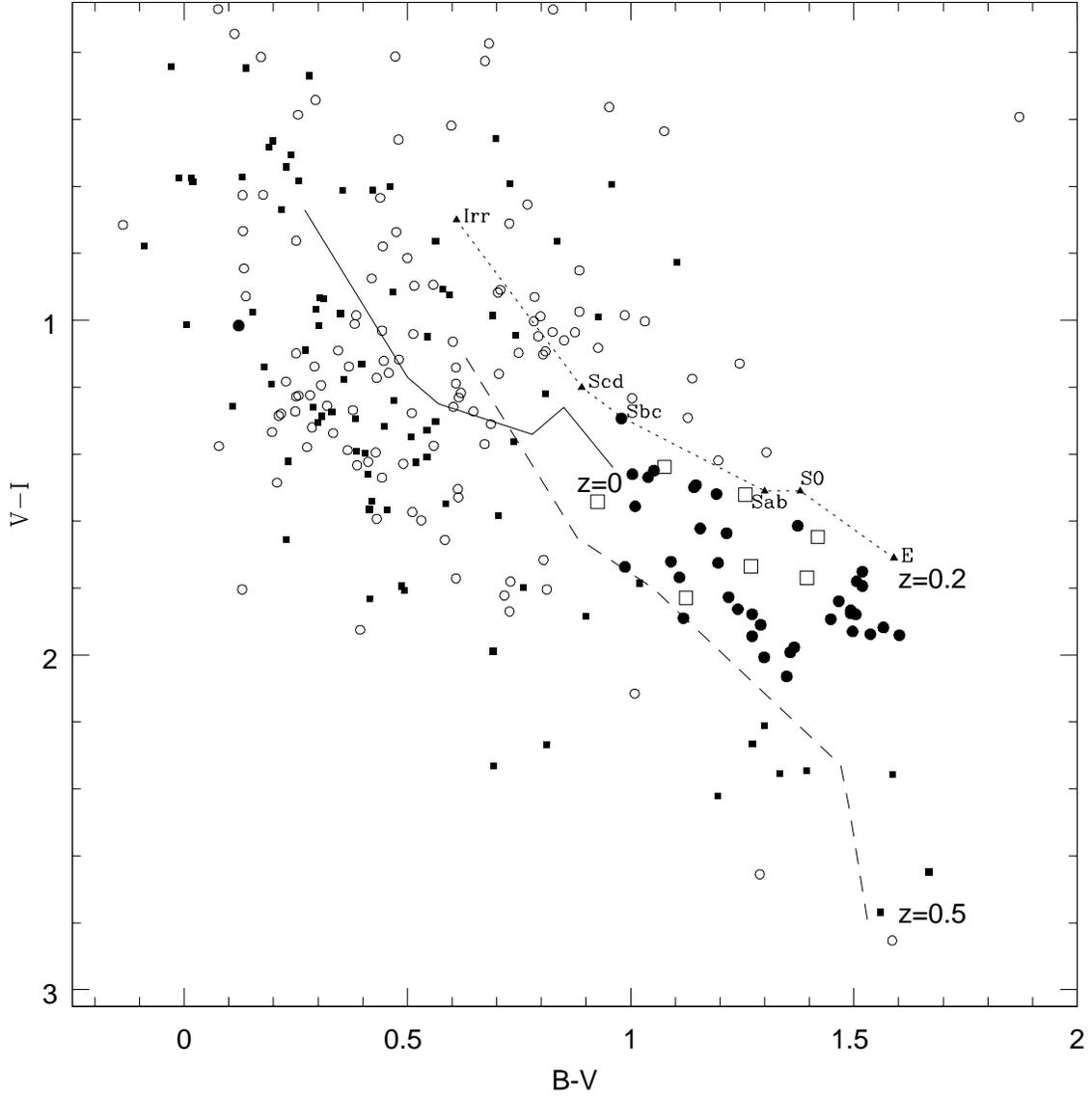}
\caption{Color-color plot of all extended
objects located within a 1\farcm 5 Mpc projected radius of the central galaxy
(circles) and all extended objects in a region of equal geometric area
elsewhere in the image (squares).
Lines connect model galaxy colors at a given redshift as calculated by
Fukugita et al. (1995) for $z=0$ (solid line), $z=0.2$ (dotted line),
and $z=0.5$ (dashed line).  The labels and triangles along the $z=0.2$
line indicate the approximate morphological type along the track.  Filled
circles are the objects selected as candidate cluster members
based on morphology and color, and open squares indicate galaxies in
the control region that would have been selected as candidate members.
The filled circle in the upper left of 
the diagram is the BCG.\label{ccplot}}
\end{figure}

The selection of candidate cluster members was performed by
analysis of all 174 SExtractor-selected galaxies within a projected 1.5 Mpc
(170\arcsec ) radius of the BCG.  A plot of the \bv, \vi ~color-color diagram
reveals an obvious grouping of objects near \bv$=1.5$, \vi $=
1.9$, as seen in Fig.~\ref{ccplot}.  A comparison with the modeled galaxy
colors of \citet{fukugita95}, shown in Fig.~\ref{ccplot} as lines
connecting different galaxy morphologies at the same redshift, 
suggests that these galaxies are at redshifts $0.2\lesssim z \lesssim 0.5$. 
Thirty-seven galaxies residing in the color-color plane between the $z=0.2$ and
$z=0.5$ model lines and with \bv$\gtrsim 1.1$, corresponding roughly to
galaxies of type Sbc and earlier, are selected as candidate cluster
members and indicated in Fig.~\ref{ccplot} by filled circles.  
Their astrometric and photometric
properties are listed in Table~\ref{candidates}, and the objects
are labeled by number in Fig.~\ref{clusterimage}.  
Galactic extinction toward M67 is low, with 
$E(\bv)=0.03$ \citep{mont93,schlegel98}, and no reddening corrections have
been applied in the analysis.

In order to determine the level of contamination due to field galaxies
satisfying the selection criteria, we analyzed an area of sky of the
same geometric area on a portion of the image outside the projected 1.5 Mpc 
radius surrounding the BCG.  118 galaxies were detected, seven of
which would have been
selected as candidate cluster members.  These false detections were
spaced fairly randomly throughout the control region, whereas the
candidate cluster members are strongly clustered around the BCG.
We therefore estimate that, of the thirty-eight candidate cluster
members, the actual number of cluster members fulfilling our selection
criteria is $31\pm 3$, assuming Poisson statistics for the
contaminating population.  We also note that, since the BCG is near
the edge of the image, there may be a similar number of cluster
members not imaged, if the cluster is symmetric about the BCG.

The cluster richness is estimated using $N_R$, the number of galaxies
above background within a 1.5 Mpc radius
brighter than two magnitudes fainter than the third-brightest cluster member
\citep{abell58}.  There are 25 cluster candidates meeting these criteria,
while one of the false detections from the control region met the
magnitude critereon.  If we assume that we are seeing only half of the
galaxy cluster, $N_R\approx 48$, and the cluster therefore has a
richness class $R=0$ to 1. 

The BCG, listed as object 1, does not meet the photometric selection
criteria for cluster membership.  This
galaxy is surprisingly blue, with \bv$=0.1$ and $V\!-I = 1.0$ \emph{before}
the consideration of any K-corrections.  
Surface-brightness profiles were determined for the BCG using the
ellipse-fitting routines  
in IRAF. The center of the galaxy was held fixed, and the position angle
and ellipticity were permitted to vary.  Neighboring sources were
masked to reduce profile contamination. The resulting profiles were
checked by subtracting a model 
galaxy from the image.  The residuals reveal some structure in the
central $1\farcs\,1 \approx 9.5\,{\rm kpc}$ in each band.  Numerous
iterations incorporating variations in several ellipse-fitting
parameters were attempted, but 
the residuals are persistent.  The residuals are strongest in the $B$
image, with a deficit of emission just to the east of the nucleus and
an excess of emission to the west.  In the $V$ and $I$ bands, the
residuals are weaker.  Due to the limits of seeing and pixilation, it
is difficult to determine the true nature of these residuals.  Since
the residuals are strongest in $B$ and weakest in $I$, we believe that
a dust lane may be present in the central regions of this galaxy.  

\begin{figure}
\plotone{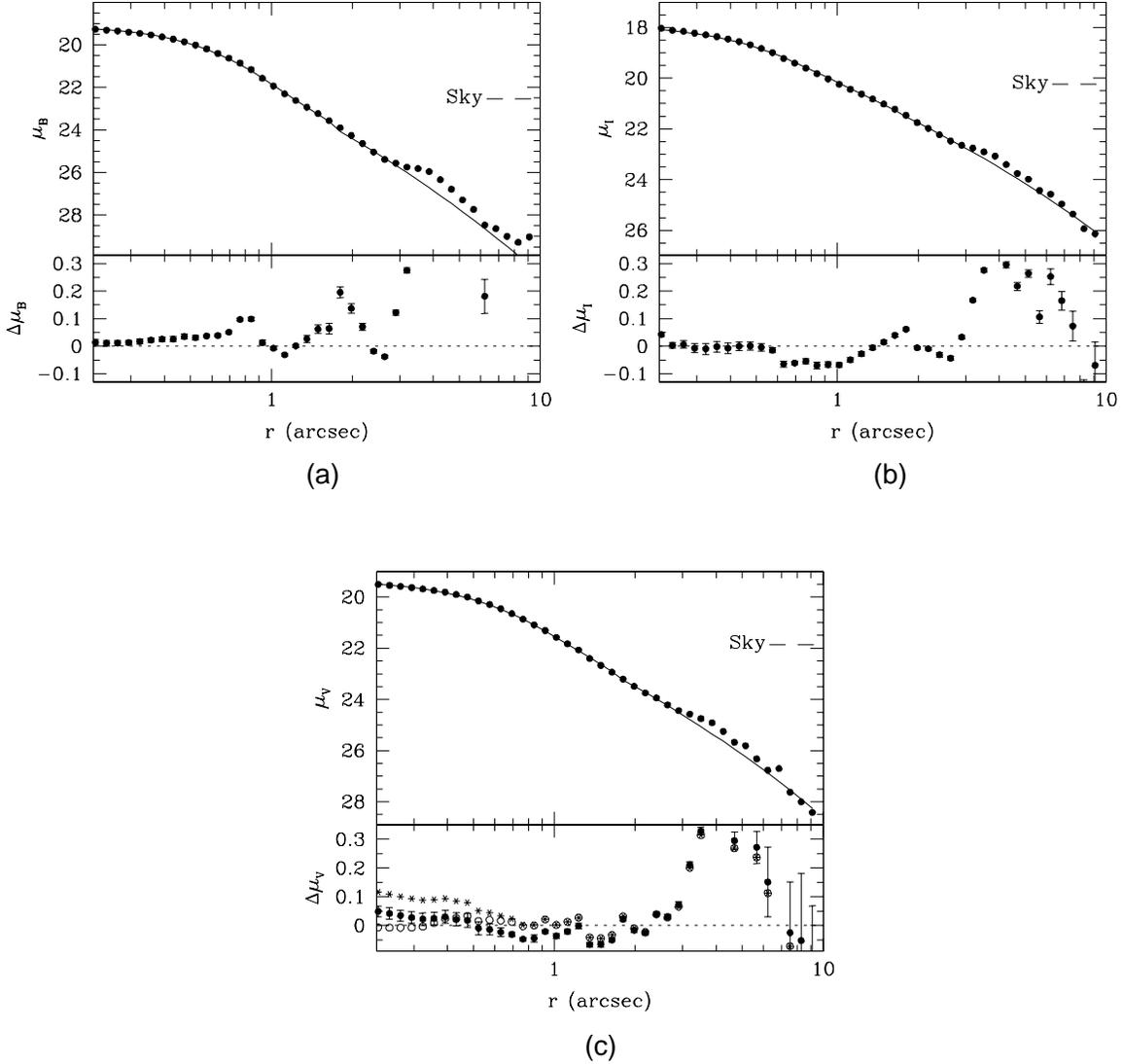}
\caption{Surface-brightness profiles and
residuals in magnitudes per square arcsecond of the BCG as
a function of radius in $B$ (a), $I$ (b), and $V$ (c) filters.  Upper
panels indicate the surface-brightness profiles; lower
panels are the residuals to the best-fitting pure $r^{1/4}$-law model
in each band.   
Solid points with error bars indicate profiles and errors obtained 
through ellipse fitting.  Solid lines are the best-fit profile
determined by the convolution of an $r^{1/4}$-law profile 
with a Gaussian PSF. The dashed lines at the right-hand side of each
upper panel indicate the measured sky brightness. Open circles in (c)
indicate the residuals to a best-fitting $r^{1/4}$-law profile with a
central point source; asterisks indicate the residuals
for the same $r^{1/4}$-law profile without a central point
source. \label{profiles}}  
\end{figure}

The surface-brightness profiles are shown in Fig.~\ref{profiles}.  The
profiles follow an $r^{1/4}$-law profile \citep{devau48} for $0\,\farcs 8
\lesssim r \lesssim 3\arcsec$.  Interior to this, seeing effects
dominate the profile.  Exterior to 3$\arcsec$, there is excess light
above the $r^{1/4}$-law profile, an effect observed in cD
galaxies.  The deviation from the $r^{1/4}$-law profile
occurs at the same radius in all three bands, though the amplitude of
the deviation varies from band to band.  This excess luminosity does
not follow the canonical cD envelope luminosity profile determined by
\citet{schombert88}: $I\propto r^{\alpha}$, with $\alpha=-1.6$.  The
measured luminosity profile is best fit with $\alpha_B = -4.1\pm 0.2$,
$\alpha_V = -3.7\pm 0.2$, and $\alpha_I = -3.1\pm 0.1$.  However, the
$\chi^2$ values are not significantly worsened by assuming an $r^{1/4}$-law
profile for the envelope luminosity profile.

The effective radius for the observed $r^{1/4}$-law profile interior to
$3\arcsec$ is
calculated by minimizing the $\chi^2$ fit of the observed profile to a
model profile comprised of an $r^{1/4}$-law profile convolved with a
Gaussian PSF and a central point source modeled by a scaled Gaussian
PSF.  The resulting fits for each band are listed in
Table~\ref{profile_tab}, and 
are plotted, along with residuals, in Fig.~\ref{profiles}.  A
Gaussian PSF was selected because the convolution of an $r^{1/4}$-law
profile with a Gaussian kernal is an analytic function.  Actual PSFs
for each band had been constructed for the M67 stellar analysis using
DAOPHOT \citep{stetson87}, and
these were compared with Gaussian functions.  Inside of one FWHM, the
difference between the actual PSFs and a Gaussian of the same FWHM
was less than a few percent, dropping to about 1\% at the core.  The
wings of the PSFs are markedly different from a Gaussian profile;
however, since the point source is presumably located at the peak of
the galaxy light 
profile, the wings have little effect on the solution.  

\begin{figure}
\plotone{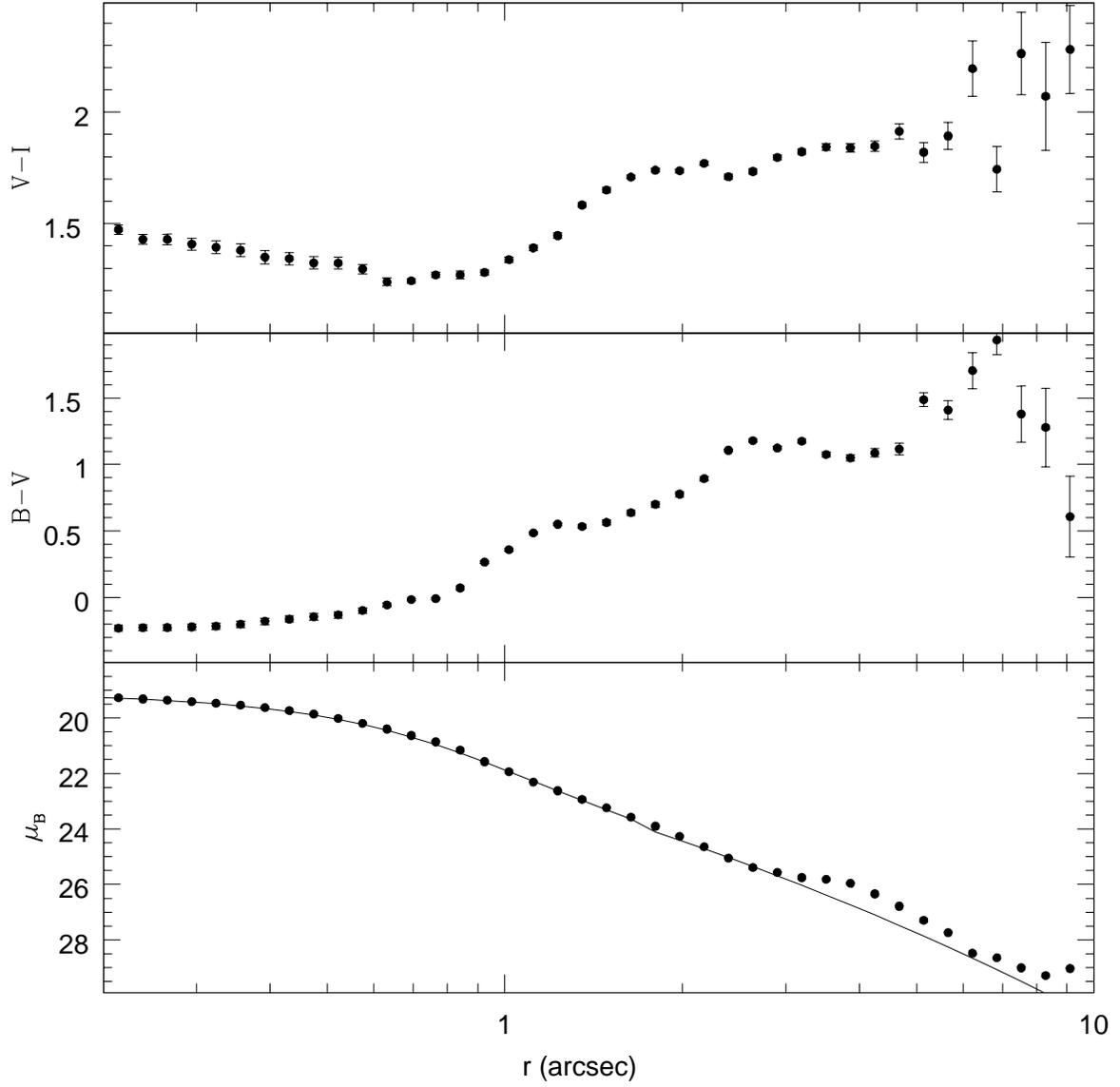}
\caption{Color as a function of radius in the central
galaxy.  The $B$-band luminosity profile is included in the bottom
panel for reference. \bv, \vi, and $\mu_B$ ~are in
units of magnitudes per square arcsecond. \label{colorprofile}}
\end{figure}

The effective radii are significantly different in each
band. As a result, the $B$-band $r^{1/4}$-law profile is dramatically steeper
than the $r^{1/4}$-law profiles in the $V$- and $I$-bands.  
The color profiles reflect this difference in effective
radii, as the galaxy becomes steadily redder with increasing radius
in both \bv and \vi, as seen in Fig.~\ref{colorprofile}.  The
colors stabilize at $\bv\approx 1.1$ and \vi$\approx 1.7$, close to the
typical colors for other early-type galaxies in this cluster. 
These color gradients are in the opposite sense of color gradients due
to metallicity gradients observed in normal
ellipticals \citep{franx89} and are similar to the gradients observed
in elliptical galaxies with blue cores in the Hubble Deep Field
\citep{menanteau00}.   

The $B$- and $I$-bands can be well described by an $r^{1/4}$-law profile.
The $V$-band
model, though, is marginally improved by the inclusion of a central
point source with $V=19.35$.  This is shown in the lower panel of
Fig.~\ref{profiles}(c), which plots the residuals of the analytic
profile compared to 
the actual surface-brightness profile.  Filled circles indicate the
best-fitting profile without a central point source.  Open circles
indicate the overall best fit, including a central point source,
whereas asterisks the same $r^{1/4}$-law profile without a central
point source.  

The presence of a point source in $V$ but not in $B$ and $I$ can be
explained in two ways.  The spectroscopic evidence presented below
suggests that there may be an AGN in the center of the galaxy.
Assuming that the AGN continuum light is blue, then the point source
should be brighter in the $V$-band
than in the $I$-band.  This explanation would
require obscuration of the AGN in $B$ to explain the non-detection of
the source in that band. An alternative
explanation is that the point source detection in $V$ is spurious.
The residuals in Fig.~\ref{profiles}(c) are smaller, but not significantly so,
for the $r^{1/4}$-law plus point source fit than for the
$r^{1/4}$-law-only fit.  Given the available data, it is not possible
to make a definitive statement about the presence or absence of a point
source in the center of the BCG.

\begin{figure}
\plotone{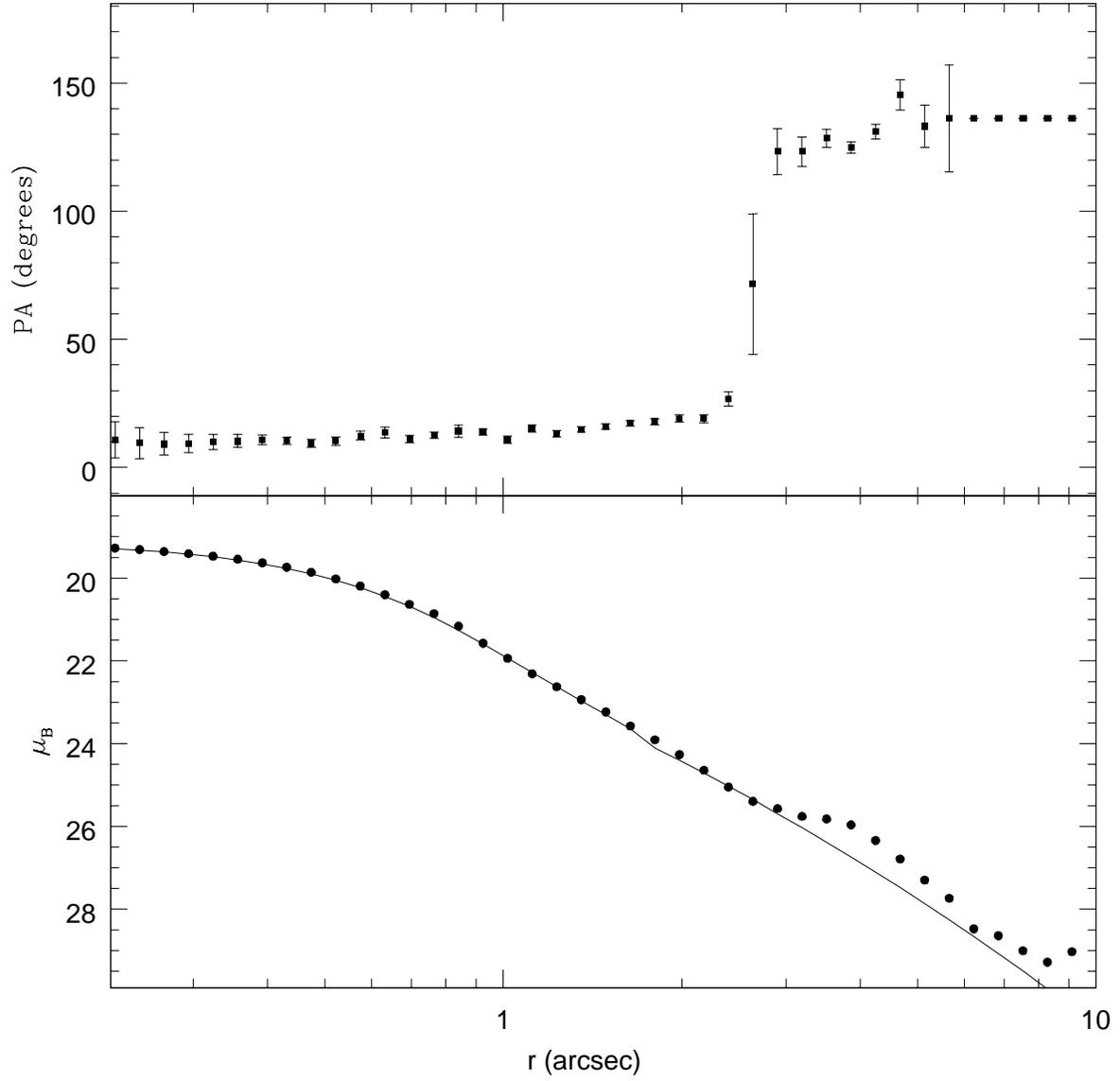}
\caption{Position angle of the BCG isophotes as
a function of radius, with the $B$-band surface-brightness profile
provided for reference. \label{pa}}
\end{figure}

The ellipticity of the isophotes does not vary significantly with
radius or bandpass,
with $e= 0.15\pm 0.05$.  The position angle is constant
with radius to $r=3\arcsec$, with a sharp $70\degr$ twist at that
point. This is the 
same radius at which the extended emission envelope appears, as shown in
Fig~\ref{pa}.  The position angle of this envelope is roughly aligned
with the 
majority of neighboring cluster member galaxies, as shown by the arrow in
Fig.~\ref{clusterimage}.  This alignment is common for cD envelopes
\citep{carter80}, and an isophotal twist at the same radius as the
start of the envelope is also not uncommon \citep{porter91}, 
providing additional evidence that the envelope is closely related to
classical cD envelopes. 

\section{Spectroscopic Results}

Spectra of candidate member galaxies were obtained as part of a
follow-up multi-slit spectroscopy of \objectname{M67} white dwarf
candidates. Only four slits were devoted to candidate
member spectroscopy due to spatial constraints on the slit masks.  One
candidate cluster member, object 10, was detected, but the
signal-to-noise was too poor to permit spectral identification. 
Three cluster galaxies, objects 1, 23, and 31,  were detected 
with sufficient signal-to-noise to permit redshift determinations.
The three redshifts are included in Table~\ref{candidates}.  

The galaxy cluster redshift and velocity
dispersion were calculated following the suggestions of
\citet{beers90} using the ROSTAT program graciously made
available by the authors.  For tiny sample sizes (five
galaxies), they determined
that the median redshift is a satisfactory measurement for numerous
distributions, with a decent estimate for the confidence interval
given by the canonical $IC_{\mu,\sigma}$ statistic.  \citet{beers90}
also state that the canonical standard deviation $\sigma$ provides a 
satisfactory estimate of the cluster velocity dispersion for tiny
samples, with confidence intervals for the velocity dispersion
adequately estimated by the jackknifed biweight. These
calculations result in a cluster redshift of $z=0.369\pm 0.003$ and
a velocity dispersion of $\sigma=290\pm 244\,{\rm km}\,{\rm s}^{-1}$,
with both errors being 68\% confidence limits. 
\citet{zabludoff98} note that
velocity dispersions of groups determined from a small number of
bright galaxies tend to be underestimated due to insufficient sampling
of the velocity distribution; this effect may likely be present in
this tiny sample.  A larger sample of redshifts is clearly necessary
before the velocity dispersion can be reasonably well determined.

\begin{figure}
\plotone{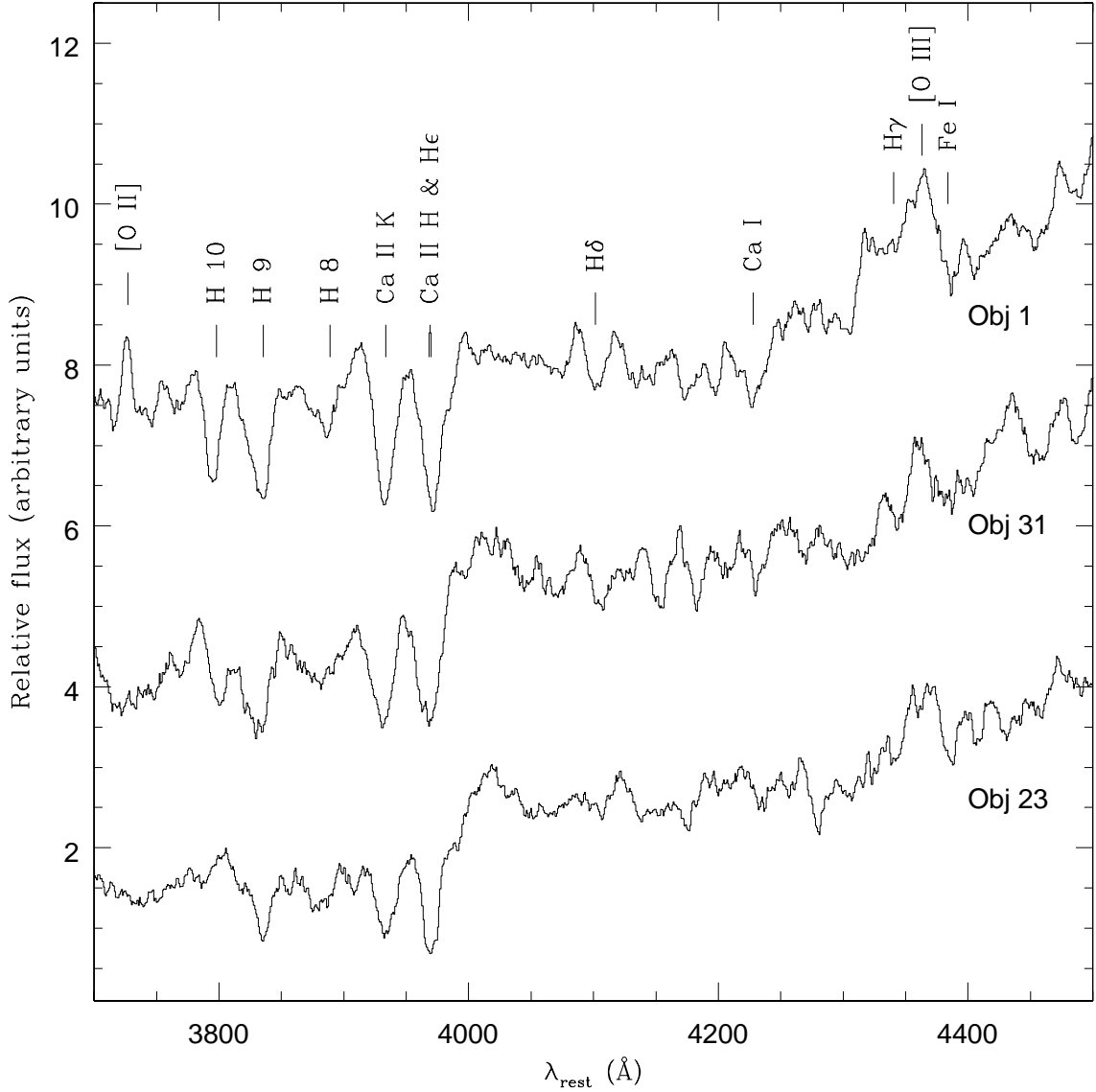}
\caption{Spectra obtained for three cluster members,
shifted to rest wavelength and boxcar smoothed by 15 pixels ($\sim
18.5$\AA).  Spectra are relatively flux-calibrated for $\lambda_{observed}\leq
4350$\AA ~and are offset vertically by arbitrary amounts for
clarity. Identified  
spectral lines are labeled. \label{spectra}} 
\end{figure}

The three spectra, smoothed and trimmed, are shown in Fig.~\ref{spectra}.   
Object 23 has a typical early-type galaxy spectrum.  The
dominant lines are the \ion{Ca}{2} H and K lines, and the 4000\AA ~break is
prominent.  The G band is 
contaminated by the \ion{Na}{1} D night-sky lines and was trimmed from
the spectrum.  Object 31 contains a mixture of a relatively
young and an older stellar population, with prominent Balmer
absorption lines along with strong \ion{Ca}{2} and \ion{Fe}{1}
absorption.  The lack of  
[\ion{O}{2}] emission at $\lambda_{rest} = 3727\,$\AA ~indicates
a lack of current star formain in these galaxies. 

\subsection{The central galaxy}
The spectrum of the central galaxy is peculiar.  As seen in
Fig.~\ref{spectra}, Balmer absorption is present at H$\delta$ and
higher-order lines, and the blended [\ion{O}{2}] emission lines have
an  equivalent width of $\sim 2$\AA ~(corrected for redshift).  The
4000\AA ~break is small.  An excess of emission is visible from
$\lambda_{rest} \approx 4300 \mbox{\AA}$ to 4400\AA  that is not
observed in the other two objects.  This emission feature is roughly
centered on H$\gamma$, which is otherwise not observed. Possible
emission features also bracket the H$\delta$ line, emission features also
not observed in the spectra of the other two galaxies.

\begin{figure}
\plotone{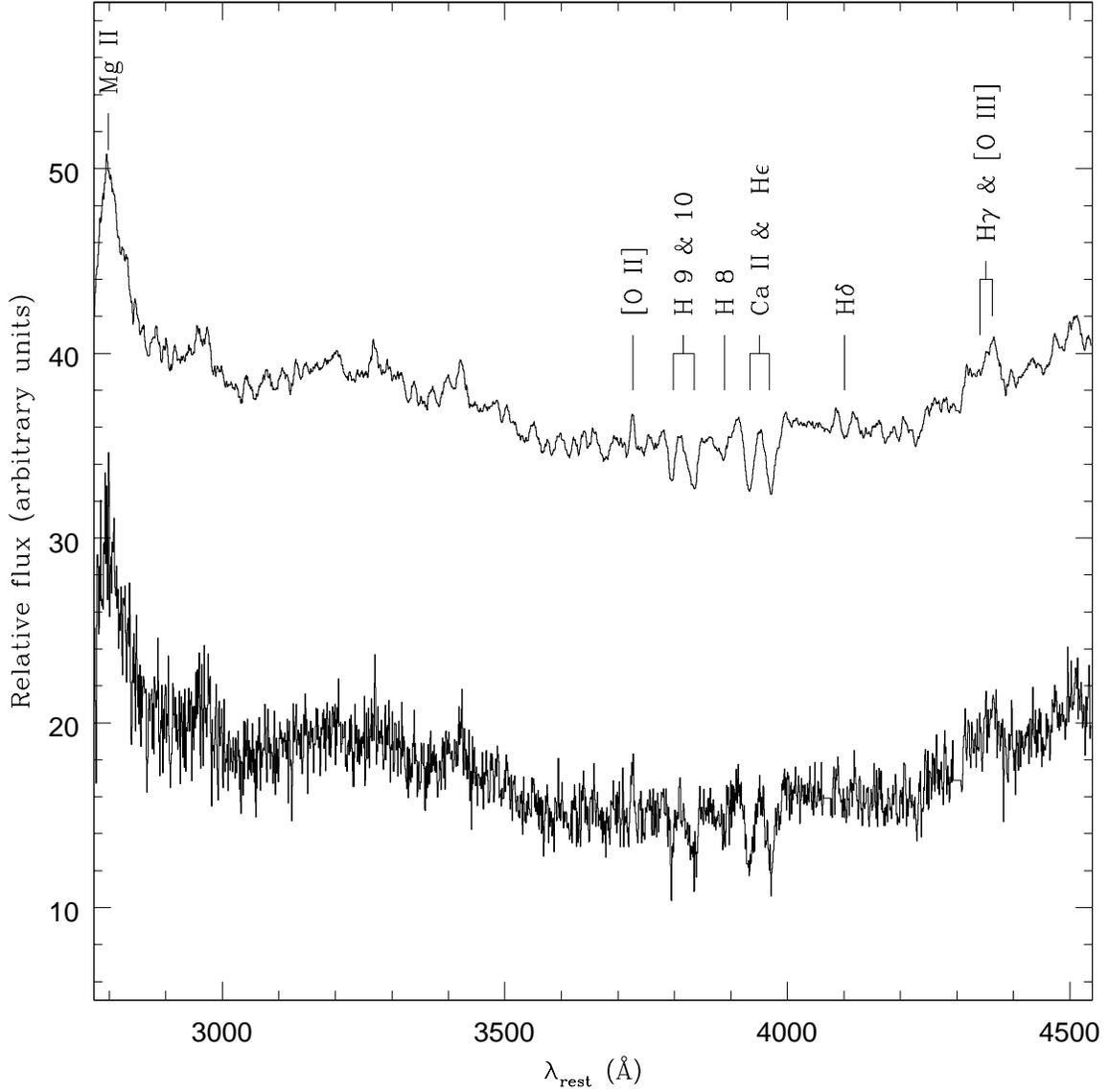}
\caption{Spectrum of the BCG at
instrumental resolution (lower) and boxcar smoothed by 11 pixels
(upper).  The upper spectrum is offset vertically by an arbitrary amount.
Relative flux calibrations are reliable for
$2850\mbox{\AA}\leq\lambda_{rest}\leq 4350\mbox{\AA}$. The
spectrum has been blueshifted to rest wavelength, and
poorly-subtracted night sky lines have been removed.
\label{cdspectrum}} 

\end{figure}

The full spectrum, shown in Fig.~\ref{cdspectrum},  is even more
interesting. A broad \ion{Mg}{2} emission line
($\lambda_{rest} = 2798 \mbox{\AA}$)
is present on the blue end of the spectrum, and the continuum flux rises
significantly blueward of $\lambda_{rest}\approx 3600\mbox{\AA}$.  
The \ion{Mg}{2} emission line has a rest-frame FWHM of $\approx
49 \mbox{\AA} = 5250 \,{\rm km}\,{\rm s}^{-1}$, suggesting the
presence of a central AGN.  This may 
explain the peculiar emission feature at
H$\gamma$, if the core of a broad emission line is filling in the
H$\gamma$ absorption line from the young stellar population.

The presence of [\ion{O}{2}] emission and strong Balmer absorption is
indicative of recent, perhaps even ongoing, star formation in the
BCG.  However, the possible contamination of the H$\gamma$ and H$\delta$
absorption lines by emission makes it difficult to measure equivalent
widths with the data currently available. Observations of H$\beta$ and
[\ion{O}{3}], which are not currently 
available, should be able to determine the degree to which
[\ion{O}{2}] is contaminated by emission from the suspected AGN, and
thereby help to determine the current star formation rate, if any
\citep{kennicutt92}. 

\section{Discussion and conclusions}

The observations of the central galaxy indicate that this BCG is a
very peculiar object.  The presence of a broad \ion{Mg}{2} emission
line suggests the presence of a central AGN.  Marginal detection of a
central unresolved source in the $V$-band photometry, as well as
possible emission wings on either side of the H$\delta$ and H$\gamma$
lines offer weak supporting evidence for an AGN.  Higher-resolution
imaging and spectroscopy at longer wavelengths (especially near
H$\beta$) should be capable of confirming any active nucleus.

The anomalously blue colors of the central regions of this galaxy,
combined with the strong Balmer absorption lines observed in the
spectrum, suggest that recent star formation has occurred in this galaxy.
Given the small equivalent width of [\ion{O}{2}] $\lambda\lambda 3727
\mbox{\AA}$, it
appears that the star formation rate in the BCG is currently low.

A scenario that may explain the observations would involve the
accretion of a gas-rich galaxy by the central elliptical galaxy.  Such
an event would supply cold gas for star formation and could explain
the possible dust feature observed in the residuals of the ellipse
fitting.  However, the currently-available data do not permit useful
constraints to be placed on any accretion scenarios.  

Several interesting questions are raised by the observations of the
BCG. Is recent star formation the source of the strong central color
gradients?  If so, how recent and vigorous was the
activity?  Could the star formation and the central nuclear activity
be related?  More detailed spectroscopic and higher-resolution imaging
are necessary to address these questions and explain this peculiar galaxy.

The central galaxy is located 112\arcsec
~from the $V=14.2$ star \objectname{MMJ 5722} \citep{mont93} and
89\arcsec ~from the $V=17.4$ star \objectname{MMJ 5568}.  In the
future, these and other M67 stars may serve as natural guide
stars for next-generation adaptive optics observations of this galaxy,
though at the 
present these stars are located at too large of an angular distance
from the central galaxy for AO work \citep{wiz00,rigaut98}.

In summary, we report the discovery of a poor
cluster or group of galaxies  at $z=0.369$ behind the galactic
stellar cluster M67.  Rough photometric redshifts result in $31\pm3$
observed member galaxies 
brighter than $V = 24$, and redshifts from three members indicate
a velocity dispersion of $\sigma = 290\pm 244$ km s$^{-1}$.  The
central galaxy has an $r^{1/4}$-law surface-brightness profile,
though the effective radius increases substantially in redder
bandpasses.  Outside a 
radius of $r= 3\arcsec\, (\approx 26\,{\rm kpc}$), light in excess of the
fit $r^{1/4}$-law profile is observed.  A central, unresolved source appears
also observed in $V$, but only upper limits on any such source can be
made in $B$ and $I$.  The spectrum
of the BCG contains strong Balmer absorption, weak
[\ion{O}{2}] emission, and a broad \ion{Mg}{2} emission line, as well
as possible broad emission lines of H$\gamma$ and H$\delta$.
The colors and spectroscopy may be indicative of recent star formation
in the BCG, but higher-resolution imaging and further spectroscopy are
needed to explain this peculiar galaxy further.
This cluster may be a
target for future adaptive optics systems and may be useful for studying the
formation and evolution of cD galaxies in poor clusters of galaxies.

\acknowledgements
The author wishes to  thank Michael Bolte and William Mathews
for many helpful discussions on this work, and to thank the referee
for numerous helpful comments that helped greatly to improve the quality of
this paper.  The author also thanks Kim-Vy Tran
for training in the art of high-$z$ spectroscopy. KW acknowledges the
gracious support of Phyllis Wattis and the Northern California chapter
of the ARCS Foundation. This work was supported in part by NASA grant 
NAG 5-8409. This research has made use of the NASA/IPAC Extragalactic
Database (NED) which is operated by the Jet Propulsion Laboratory,
California Institute of Technology, under contract with the National
Aeronautics and Space Administration.  

\break

\clearpage

\begin{deluxetable}{crrccccccc}
\rotate
\renewcommand{\arraystretch}{.6}
\tablewidth{0pt}
\tablecaption{Candidate cluster member astrometry, photometry, and
spectroscopic redshifts.\label{candidates}} 
\tablehead{\colhead{Object} &\colhead{RA(J2000.0)}&\colhead{Dec(J2000.0)}
&\colhead{$V$} & \colhead{$\delta V$} &\colhead{\bv} & \colhead{$\delta(\bv )$}
&\colhead{$V\! -I$} & \colhead{$\delta(V\! -I)$} & \colhead{$z$}} 
\startdata
1  & 8 51 16.42&+12 \phn 0 29.4&18.77&0.10&0.12&0.13&1.02&0.13&0.3705 \\
2  & 8 51 \phn 9.89&+12 \phn 0 33.8&21.88&0.12&1.27&0.17&1.94&0.17&\nodata \\
3  & 8 51 11.01&+12 \phn 0 17.1&20.99&0.10&1.60&0.16&1.94&0.14&\nodata \\
4  & 8 51 11.08&+12 \phn 0 11.5&23.29&0.14&1.14&0.27&1.50&0.21&\nodata \\
5  & 8 51 11.38&+12 \phn 0 \phn 9.9&23.67 & 0.16&1.05&0.32&1.45&0.23&\nodata \\
6  & 8 51 11.70&+12 \phn 0 21.3&22.11&0.13&1.47&0.19&1.84&0.18&\nodata \\
7  & 8 51 11.72&+11 59 45.1&22.57&0.16&1.30&0.21&2.01&0.20&\nodata \\
8  & 8 51 12.99&+11 58 26.3&23.20&0.14&1.35&0.29&2.06&0.19&\nodata \\
9  & 8 51 13.04&+11 59 27.7&21.47&0.11&1.52&0.16&1.79&0.16&\nodata \\
10 & 8 51 13.36&+12 \phn 0 12.2&21.57&0.11&1.15&0.17&1.49&0.16&\nodata \\
11 & 8 51 13.52&+12 \phn 0 \phn 8.7&21.73&0.12&0.98&0.17&1.29&0.17&\nodata \\
12 & 8 51 13.75&+11 58 51.9&20.86&0.10&1.54&0.16&1.94&0.14&\nodata \\
13 & 8 51 13.81&+12 \phn 0 27.2&22.30&0.15&1.20&0.19&1.72&0.19&\nodata \\
14 & 8 51 15.20&+11 59 15.8&22.52&0.16&1.37&0.21&1.98&0.20&\nodata \\
15 & 8 51 15.57&+11 59 51.4&22.76&0.16&1.22&0.22&1.64&0.20&\nodata \\
16 & 8 51 15.71&+12 \phn 0 11.4&22.21&0.14&1.45&0.19&1.89&0.18&\nodata \\
17 & 8 51 15.83&+12 \phn 0 25.1&21.94&0.12&1.24&0.17&1.86&0.17&\nodata \\
18 & 8 51 16.26&+11 59 54.3&23.44&0.14&1.11&0.29&1.77&0.21&\nodata \\
19 & 8 51 16.51&+12 \phn 0 15.5&20.13&0.09&1.57&0.14&1.92&0.13&\nodata \\
20 & 8 51 16.78&+12 \phn 0 \phn 5.7&21.76&0.12&1.16&0.17&1.62&0.17&\nodata \\
21 & 8 51 16.87&+12 \phn 0 \phn 7.7&21.31&0.11&1.36&0.16&1.99&0.15&\nodata \\
22 & 8 51 16.97&+12 \phn 0 \phn 9.7&21.67&0.12&1.22&0.17&1.83&0.16&\nodata \\
23 & 8 51 17.25&+12 \phn 0 24.0&21.28&0.10&1.50&0.16&1.93&0.15&0.3685 \\
24 & 8 51 17.31&+12 \phn 0 20.0&23.10&0.14&0.99&0.22&1.74&0.20&\nodata \\
25 & 8 51 17.41&+12 \phn 0 19.2&22.69&0.16&1.09&0.21&1.72&0.20&\nodata \\
26 & 8 51 18.18&+11 59 59.5&22.95&0.15&1.37&0.26&1.61&0.20&\nodata \\
27 & 8 51 18.45&+11 58 30.3&23.41&0.14&1.51&0.32&1.78&0.21&\nodata \\
28 & 8 51 18.85&+12 \phn 0 3.2&23.94&0.20&1.04&0.35&1.47&0.27&\nodata \\
29 & 8 51 19.02&+11 59 42.9&21.32&0.11&1.50&0.16&1.88&0.15&\nodata \\
30 & 8 51 19.21&+11 59 14.6&20.54&0.10&1.49&0.16&1.87&0.14&\nodata \\
31 & 8 51 19.64&+12 \phn 0 \phn 2.7&21.78&0.12&1.12&0.17&1.89&0.17&0.3680\\
32 & 8 51 19.70&+11 59 59.6&21.39&0.11&1.27&0.17&1.88&0.15&\nodata \\
33 & 8 51 20.11&+11 59 55.7&21.74&0.12&1.49&0.17&1.88&0.17&\nodata \\
34 & 8 51 20.19&+12 \phn 0 25.8&22.93&0.15&1.01&0.21&1.56&0.21&\nodata \\
35 & 8 51 20.56&+12 \phn 0 \phn 5.4&23.23&0.14&1.00&0.24&1.46&0.21&\nodata \\
36 & 8 51 24.75&+11 59 33.3&22.30&0.15&1.52&0.20&1.75&0.19&\nodata \\
37 & 8 51 26.19&+12 \phn 0 \phn 3.9&22.21&0.14&1.29&0.19&1.91&0.18&\nodata \\
38 & 8 51 26.36&+11 59 45.4&21.94&0.13&1.19&0.18&1.52&0.17&\nodata \\
\enddata
\end{deluxetable}

\clearpage

\begin{deluxetable}{ccccc}
\tablewidth{0pt}
\tablecaption{Central galaxy photometric parameters. \label{profile_tab}}
\tablehead{\colhead{Band} & \colhead{$r_e$} & \colhead{$r_e$} &
\colhead{$\mu_e$} & $m_*$\tablenotemark{a}\\
\omit & \colhead{(\arcsec)} & \colhead{(kpc)} &
\colhead{mag arcsec$^{-2}$} & \colhead{mags}}
\startdata
$B$ & $0.349\pm 0.022$ & 3.07 & $19.87\pm 0.10$ & $\gtrsim 22$ \\
$V$ & $0.828\pm 0.044$ & 7.29 & $21.52\pm 0.06$ & $19.35\pm 0.15$ \\
$I$ & $1.308\pm 0.087$ & 11.5 & $20.87\pm 0.06$ & $\gtrsim 20$ \\
\enddata
\tablenotetext{a}{Magnitude of central point source}
\end{deluxetable}

\clearpage

\clearpage

\clearpage

\clearpage

\clearpage

\clearpage

\clearpage

\clearpage

\end{document}